\documentclass{aa}
\usepackage{graphicx}
\usepackage{epsfig}
\newcommand{\pp}{$P-\dot{P}$~}
\def\la{\mathrel{\mathchoice 
{\vcenter{\offinterlineskip\halign{\hfil$\displaystyle##$\hfil\cr<\cr\sim\cr}}}
{\vcenter{\offinterlineskip\halign{\hfil$\textstyle##$\hfil\cr<\cr\sim\cr}}}
{\vcenter{\offinterlineskip\halign{\hfil$\scriptstyle##$\hfil\cr<\cr\sim\cr}}}
{\vcenter{\offinterlineskip\halign{\hfil$\scriptscriptstyle##$\hfil\cr<\cr\sim\cr}}}}}
\def\ga{\mathrel{\mathchoice 
{\vcenter{\offinterlineskip\halign{\hfil$\displaystyle##$\hfil\cr>\cr\sim\cr}}}
{\vcenter{\offinterlineskip\halign{\hfil$\textstyle##$\hfil\cr>\cr\sim\cr}}}
{\vcenter{\offinterlineskip\halign{\hfil$\scriptstyle##$\hfil\cr>\cr\sim\cr}}}
{\vcenter{\offinterlineskip\halign{\hfil$\scriptscriptstyle##$\hfil\cr>\cr\sim\cr}}}}}
\titlerunning{Monopolar Pulsar Spin-Down}
\authorrunning{Alvarez \& Carrami\~nana}
\begin{document}
     \title{Monopolar Pulsar Spin-Down}
%
%
     \author{C\'esar Alvarez \and Alberto Carrami\~nana}
     \offprints{C\'esar Alvarez}

     \institute{Instituto Nacional de  Astrof\'{\i}sica, \'Optica y Electr\'onica,\\ 
       Luis Enrique Erro 1, Tonantzintla, Puebla 72840, M\'exico
\\
     \email{calvarez@inaoep.mx, alberto@inaoep.mx}
}
     \date{Sent 15 January 2003}

\abstract{
A multipole spin-down equation based on a monopolar term is derived from the 
general expression $\dot{\nu} = -f(\nu,t)$ and used to study pulsar evolution. 
We show that the time-independent version of such equation cannot reproduce the
observed properties of pulsars and conclude that there is no equation of the form 
$\dot{\nu} = -f(\nu)$ consistent with the \pp diagram and braking index measurements. 
We explore the time-dependent model under the hypothesis of decaying magnetic 
fields, showing that an inverse linear decay gives reasonable evolutionary 
trajectories. This model distinguishes the evolution of Vela from that of the other
three young pulsars considered. We discuss the origin of the monopolar term, which 
cannot be attributed to radiative processes, pointing to the importance of particle 
acceleration and/or mass loss processes in the dynamical evolution of pulsars.
}

\maketitle

\section{Introduction}
Pulsars are accepted to be rapidly rotating and highly magnetized neutron stars 
powered by the loss of rotational kinetic energy. Yet, the mechanisms through 
which pulsars lose their rotational energy are not very well understood. The 
discovery of pulsars was preluded by the rotating magnetic dipole model of 
Pacini (\cite{pacini67}), which predicted a spin-down relation $\dot{\nu}\propto 
\nu^{3}$, where $\nu$ is the rotational frequency. Pulsar spin-down has 
then been studied through the power-law differential equation,
\begin{equation}
\dot{\nu}=-k\nu^n \, ,
\end{equation}
where $k$ is a constant and $n$ is called the ``braking index'', determined by 
the physical mechanisms spinning-down the star. For a pure magnetic dipole $n$ is
equal to 3 (Gold \cite{gold}; Pacini \cite{pacini68}), while a pure gravitational or 
electromagnetic quadrupole one has $n=5$ (Ostriker \& Gunn \cite{ostriker};
Ferrari \& Ruffini \cite{ferrari}). The expansion of the Larmor equation into 
electromagnetic moments higher than dipolar gives $n > 3$. Departures from these basic 
models have been 
studied. For example, the deformation of magnetic field lines by corotation with the 
magnetosphere might produce $1\leq n\leq 3$ (Manchester \& Taylor \cite{manchester77}), 
pulsar winds $n<3$ (Blandford \& Romani \cite{blandford}; Manchester et al. 
\cite{manchester85}), while magnetic field decay (Chanmugam \& Sang \cite{chanmugam89})
and the alignment between rotation and magnetic axis tend to give $n>3$ 
(Goldreich \cite{goldreich}). 
{Physical processes such as phase transitions 
in the interior of the fast rotating neutron stars can produce changes in the moment of 
inertia $I$ and led also to deviations from $n=3$ even if the pulsar is spinning-down by 
pure dipolar radiation (Chubarian et al. \cite{chubarian2000}; Glendenning et al. 
\cite{glendenning97}). 
However, these phase transitions are significant only for very fast pulsars, with periods 
$P<1.5$ ms. For pulsars with periods $P\ga 3 $ ms, these changes in the moment of inertia 
can be neglected.} But while physical mechanisms to produce $n<3$ have been
proposed, 
{the reconciliation of pulsar evolution with $n<3$ remains to be studied}. 
This 
{question, addressed to the case of young classical pulsars,} is one of the main 
objectives of this work.

The braking index $n$ is frequently defined in terms of observational quantities,
with a definition generalized to a second braking index, $m$:
\begin{equation}
n \equiv {\ddot{\nu}\nu \over \dot{\nu}^2} \, , \qquad 
m\equiv {\stackrel{...}{\nu}\nu^2\over\dot{\nu}^3} \, .
\label{index}
\end{equation}
Accurate measurement of both braking indices requires long-term timing measurements, 
a difficult task due to the existence of glitches and timing noise in young pulsars, 
affecting the measurement of $\ddot{\nu}$ and $\stackrel{\ldots}{\nu}$. Out of more 
than 1300 detected pulsars, only a handfull have reliable braking index measurements 
(Table~\ref{tbl}). Still, measurements continue to be made and objects like 
PSR~J1119--6127 and PSR~J1846--0258 might probably be recent additions to the list 
(Camilo et al. \cite{camilo00}; Mereghetti et al. \cite{mereghetti}).
The second braking index $m$ is even more difficult to measure and has only been 
estimated for PSR~B1509--58~(Kaspi et al. \cite{kaspi}) and the Crab pulsar~
(Lyne et al. \cite{lyne88})
(Table~\ref{tbl}). It is well-known that measured values of $n$ and $m$ are below, 
but close, to those of a magnetic dipole in vacuum ($n=3$ and $m=15$).

\begin{table}
\caption[]{Timing parameters and stationary multipole model fitting parameters
for the four young pulsars selected.}
\label{tbl}
$$
\begin{array}{lcccc}
\hline
\noalign{\smallskip}
Pulsar & \mathrm{Crab}^{(1)} & 1509-58^{(2)} & 0540-69^{(3)} & \mathrm{Vela}^{(4)}
\\
\noalign{\smallskip}
\hline
\noalign{\smallskip}
P ($ms$) & 33.5  & 150.9 & 50.3 & 89.3 \\
t_{dyn} ($yr$)  &  1258 & 1553 & 1664 & 11300 \\
n & 2.509 & 2.837 & 2.01 & 1.4 \\
  & \pm 0.005 & \pm 0.001 & \pm 0.02 & \pm 0.2 \\
m & 10.23 & 14.5 & \ldots & \ldots \\
 & \pm 0.03 & \pm 3.6 & \ldots & \ldots \\
\hline
\noalign{\smallskip}
g ($Hz$^{-3}) & 3.5\cdot 10^{-19} & 6.2\cdot 10^{-16} & 6.6\cdot 10^{-18} & 3.8\cdot 10^{-18} \\
r ($Hz$^{-1}) & 1.0\cdot 10^{-14} & 1.6\cdot 10^{-13} & 7.1\cdot 10^{-15} & 1.3\cdot 10^{-15} \\
s ($Hz$) & 3.4\cdot 10^{-12}  & 2.0\cdot 10^{-12} & 5.7\cdot 10^{-12} & 1.2\cdot 10^{-12} \\
\noalign{\smallskip}
\hline
P_{birth} & 9.9 & 14.2^{b} & 17.6 & 50 \\
$(ms)$^{a} &  &  &  & \\
\hline
\end{array} 
$$
\begin{list}{}{}
\item[$^{\mathrm{a}}$]$P_{birth}$ defined as the period one dynamical time ago.
\item[$^{\mathrm{b}}$]Integration is stopped at $t_{dyn}$ = 1~year.
(1) Lyne et al. (\cite{lyne88}); (2)~Kaspi et al. (\cite{kaspi}); (3)~Manchester
\& Peterson (\cite{manchester89}); (4)~Lyne et al. (\cite{lyne96}).
\end{list}
\end{table}

Pulsar evolution is mostly studied through the \pp diagram, which shows a sparse 
distribution of pulsars clustered roughly at $\log P\sim -0.3$ and $\log\dot{P}
\sim -15$. In principle it should be possible to link the bulk of the pulsar population 
in the \pp diagram with the younger pulsars through evolutionary tracks consistent with a 
general spin-down equation. However, neither the magnetic dipole model or a constant 
braking index equation can link these groups and it has been argued that pulsars might 
be born with small period derivatives, in a region of the \pp diagram where no pulsar 
detection has been reported (Camilo \cite{camilo96}). Alternatively, magnetic dipole 
moments might decay during the pulsar life (Camilo \cite{camilo96}; Tauris \& Konar 
\cite{Tauris}). 
On these lines, {Colpi et al.~(\cite{colpi2000}) studied the distribution and 
evolution in the \pp diagram of anomalous X-ray pulsars (AXPs), considering 
dipolar spin-down of magnetars. They conclude that magnetic field decay is required 
to account
for the observed distribution and X-ray luminosities.}

We propose here a spin-down model derived from the very general spin-down law 
$\dot{\nu}=-f(\nu,t)$. Although the derivation is empirical, the model accomodates
the usual spin-down mechanisms, namely magnetic dipole radiation and gravitational 
radiation. In addition, a monopolar term is introduced, aiming to reconcile braking
indices with trajectories in the \pp diagram. We calculate evolutionary trajectories 
for PSR~B1509--58, PSR~B0540--69, the Vela pulsar (PSR~B0833--45) and the Crab pulsar  
(PSR~B0531+21). The model is presented in Section~2, while in Section~3 we apply the 
model to the four pulsars just mentioned, assuming time-independent spin-down mechanisms.
In Section~4 we study the model considering time-dependent spin-down mechanisms. 
In Section~5, a discussion of possible physical mechanisms 
related to the spin-down terms is given, to lead to the 
concluding Section~6.

\section{The Model}
The basic assumption of this work is that the frequency derivative $\dot{\nu}$ can 
be described as a function of frequency $\nu$ and time $t$ only:
\begin{equation}
\dot{\nu}=-f(\nu,t) \, .  
\label{general}
\end{equation}
We then assume the following properties: 
\begin{enumerate}
\item $f>0$ for all $(\nu,t)$, as $f$ describes energy and angular momentum losses.
\item $f$ is a continuous function of time. This requires that the cumulative effect 
of glitches on evolutionary paths in the \pp diagram can be neglected.
\item $f$ is an 
antisymmetric function, $f(-\nu,t)=-f(\nu,t)$. This is equivalent 
to ascribe a sign to the frequency $\nu$ and its derivative $\dot{\nu}$: $\nu$
and $\dot{\nu}$ have the same signs when the pulsar is spinning-up and opposite 
signs when the pulsar spins-down.
This conditions will be further discussed below.
\end{enumerate}
The model consists then of a simple Taylor expansion of $f$, restricted to the three 
lowest
order terms:
\begin{equation}
\dot{\nu}=-s(t)\nu-r(t)\nu^3-g(t)\nu^5 \, .   \label{model}
\end{equation}   
Higher order terms can be neglected for the frequencies measured in known isolated 
pulsars\footnote{Binary and millisecond pulsars are not considered here.}. 
{A variable moment of inertia, $\dot{I}\neq 0$, would introduce an extra 
positive monopole term $\dot{I}/2I$ that can be incorporated in $s(t)$. 
This would affect only very fast
pulsars and we will consider $I$ constant in here.}
Equation~(\ref{model}) represents a very general spin-down model for isolated pulsars. 
Beyond its simple empirical derivation, it can represent the loss of rotational energy 
through physical processes described by standard electromagnetic and gravitational
radiation multipolar terms, together with the ad-hoc ``monopolar'' term. We will 
argue
later that this term has to be associated with other type of processes, 
like particle acceleration or pulsar winds. We underline that if this model cannot 
be used to interpret the \pp diagram, then there is no analytical equation of the 
form given by equation~(\ref{general}), consistent with the required conditions, 
able to reconcile braking indices with pulsar evolution.

\section{The stationary multipole spin-down model}
\begin{figure}
\centering
\epsfig{file=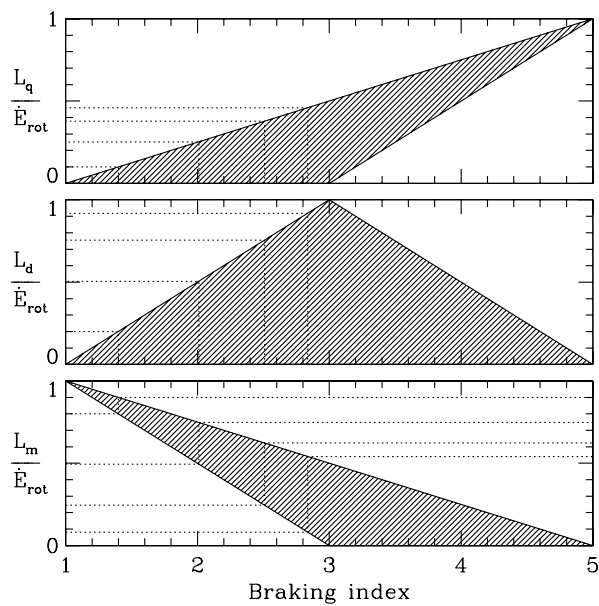, width=0.95\hsize}
\caption{$L/\dot{E}_{rot}$ ratios for the quadrupole ({\em upper panel}), dipole 
({\em middle panel}) and monopole ({\em lower panel}) terms. Only values inside the 
shaded areas are allowed by the stationary version of the model. When $n\to 1$ and
$n\to 5$ the rotational losses are due only to the monopole or quadrupole term 
respectively. The observed braking indices give upper and lower limits for the different
$L/\dot{E}_{rot}$ ratios, as indicated by the dotted lines for Vela, PSR~B0540--69, 
the Crab and PSR~B1509--58 (from left to right in the horizontal axis).}
\label{ef}
\end{figure}

\begin{figure}
\centering
\epsfig{file=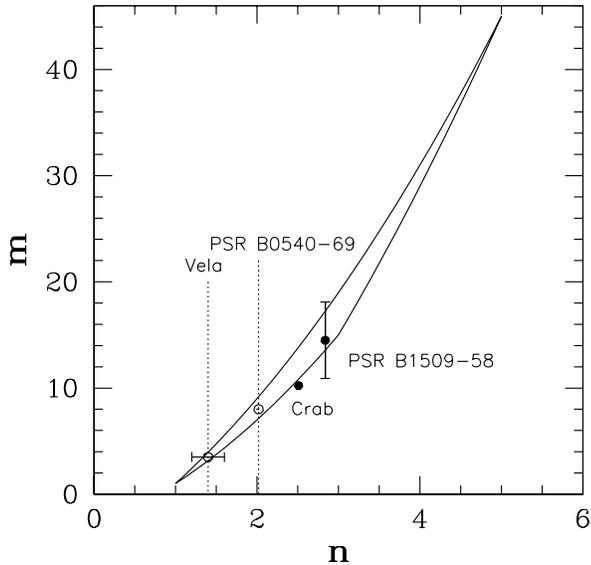, width=0.95\hsize}
\caption{Allowed values of the braking indices $n$ and $m$. According to the stationary
version of the model, pulsars evolve from the upper-right to the lower-left, increasing
their braking indices, but keeping them inside the triangle-like region. Both indices 
have been measured only for PSR~B1509--58, which lies inside the allowed region, and 
the Crab pulsar, located slightly outside the allowed region. For Vela and 
PSR~B0540--69 the value of $m$ has not been measured and we have assumed $m$ to be
in the middle of the allowed range (open circles). Vertical dot lines indicate the 
value of $n$ of these pulsars. Errors of $n$ for the Crab and PSR~B1509-58 are 
smaller than the dots.}
\label{indices}
\end{figure}

Leaving momentarily aside the discussion of the origin of the different terms, we 
consider now the model given by equation~(\ref{model}) when the coefficients $g, r$ 
and $s$ are constant, following Alvarez \& Carrami\~nana (\cite{alvarez98}). 
These coefficients can be calculated if the frequency and both braking indices are 
known. Taking the first and second derivative of equation~(\ref{model}) and using 
equations~(\ref{index}), one arrives to a matrix equation which can be inverted to 
give:
\begin{equation}
\left( \begin{array}{c} 
g \nu^5/(-\dot{\nu}) \\ r\nu^3/(-\dot{\nu}) \\ s\nu~/(-\dot{\nu})
\end{array} \right ) = { 1\over 8} 
\left( \begin{array}{rrr} 3 & -3   & 1  \\ -10  & 10 & -2 \\  15 & -7 & 1
\end{array} \right ) \left( \begin{array}{c} 1 \\ n \\ m-n^2 
\end{array}\right ) \, .
\label{matrix}
\end{equation}
The factors $g\nu^5/(-\dot{\nu})$, $r\nu^3/(-\dot{\nu})$ and $s\nu/(-\dot{\nu})$ 
represent the fractions of the total energy loss contained in the quadrupole, dipole 
and monopole terms respectively. Constraining each term to be between 0 and 1 leads 
to the bounds on their values as function of $n$ represented in Figure~\ref{ef}. 
The value of $m$ is restricted to the range
\begin{equation}
{\rm Max}(n^2+3n-3;~ n^2+7n-15)\leq m\leq n^2+5n-5\, ,   \label{mrange}
\end{equation} 
while $1\leq n\leq 5$. The minimum and maximum allowable values for $m$ are 1 and 45, 
which happen for $n=1$ and $n$=5 respectively.  Accordingly, pulsars must be in a 
restricted region of the $(n,m)$ plane defined by~equation~(\ref{mrange}), 
and shown in Figure.~\ref{indices}. Spin-down evolution is initially dominated by the 
quadrupole term ($n\to 5$), afterwards by the dipole term ($n\to 3$) and finally by 
the monopole term ($n\to 1$). Under this model the braking index itself would be a
rough qualitative age indicator.


\begin{figure}
\centering
\epsfig{file=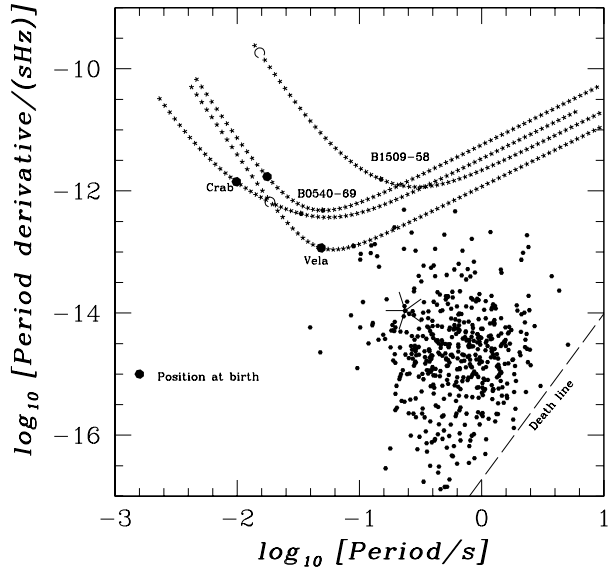, width=0.95\hsize}
\caption{Evolutionary trajectories computed for the Crab, PSR~B1509--58, PSR~B0540--69 
and Vela. The small dots on the tracks mark the present positions of these four pulsars, 
while the big dots mark the positions integrated back one dynamical time. The open 
circle marks the position of the Crab at the time of the SN1054 explosion. For 
PSR~B1509--58 the integration converges before one dynamical time, the large filled 
circle marking the \pp position with $t_{dyn}=1~\rm year$. The stationary model takes 
these pulsars into a region of the \pp diagram where pulsars have not been detected.}
\label{tr1}
\end{figure}

\begin{figure}
\epsfig{file=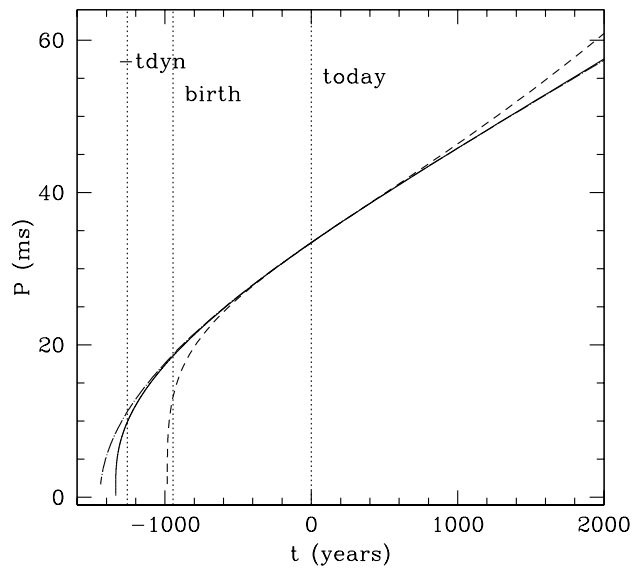, width=0.95\hsize}
\caption{Pulse period as function of time predicted for past and near-future evolution 
of the Crab pulsar. The evolutionary trajectories are given by the all allowed values 
of $m$. The dashed, full and dot-dashed lines correspond to $m=$13.64, $m=$11.0 and 
$m=$10.82 respectively. The vertical dotted lines mark the present time ($t=0$), the 
moment of the SN1054 explosion ($t=-948$~yrs) and one dynamical age backwards.}
\label{p0crab}
\end{figure}

From the known timing parameters,  $\nu$, $\dot{\nu}$, $n$ and $m$, one determines 
$\{ g, r, s\}$ uniquely, from the matrix relation~(\ref{matrix}), with no degree of 
freedom. This could only be done for PSR~B1509--58, while for the Crab, 
PSR~B0540--69 and Vela pulsars we assumed $m$ within
the constraints of the model. Evolutionary tracks were calculated for these pulsars, 
integrating  equation~(\ref{model}) backwards one dynamical time and forward to a 
point where the monopole term dominates the spin-down evolution (Fig.~\ref{tr1}).
The results for each of the four pulsars can be summarized as follows:
\begin{itemize}
\item the second braking index of the {\bf Crab pulsar}, \mbox{$m=10.23$,} is below the 
minimum value
allowed by the model ($m\geq 10.82$) and in order to compute
evolutionary trajectories we used $m=11$. This value of $m$ is consistent with the 
initial period calculated in other models for this pulsar~(Mereghetti et al. 
\cite{mereghetti}; Glendenning \cite{Glendenning}). The Crab is the only pulsar 
studied here with the precise age known, allowing some specific comparisons. 
Integrating back to its actual birth (949~years ago) one finds an initial period  
$\la 18~\rm ms$ (Fig~\ref{p0crab}). If we integrate back one dynamical time 
(1258~years) from now, we get an initial period of 10~ms, lower but within a factor 
of two of the period at birth.
\item for {\bf PSR~B1509--58} the measured values of $m$ and $n$ are consistent with 
the model. However, the integration of equation~(\ref{model}) could not be carried out 
one entire dynamical time backwards, as $\int dt$ converges at about 1350~years, 
i.e., 
before the present dynamical time (1553~years). Integrating back to a dynamical 
age of 1~year, we obtain a putative ``initial period'' of about 14~ms and a $\dot{P}$ 
two 
orders of magnitude higher than for any of the other three pulsars. This pulsar 
seems 
to have been formed in a particular region of the \pp diagram. Its quadrupolar 
parameter $g$ is at least two orders of magnitude higher that for the other pulsars.
An initial phase of strong gravitational spin-down would naturally lead to the high 
$P$ and $\dot{P}$ observed in this pulsar.
\item {\bf PSR~B0540--69:} its second braking index is still unknown so we took $m=8.0$, 
in the middle of the range of allowed values.  As the known timing parameters of 
PSR~B0540--69 are similar to those of the Crab, albeit a somewhat longer period, 
their evolutionary paths are similar, with an initial \pp position 
of PSR~B0540--69 relatively close to that of the Crab.
\item {\bf Vela pulsar:} the frequent glitches of Vela prevent any estimate of 
the second braking index, and render difficult measuring its first braking index. 
We assumed 
$m=3.5$, consistent with the low first braking index, $n=1.4\pm0.2$ 
(Lyne et al. \cite{lyne96}). As $n<2$, Vela has already passed its minimum $\dot{P}$ 
value and its predicted evolution is close to a $n=1$ line. We note that the track 
computed for the past evolution of Vela is close of that of PSR~B0540--69. However, 
its motion in the \pp diagram is very slow and when integrating back one dynamical 
time Vela only reaches $P\approx 50~\rm ms$ with a braking index almost equal to 2.
\end{itemize}

\begin{figure}
\centering
\epsfig{file=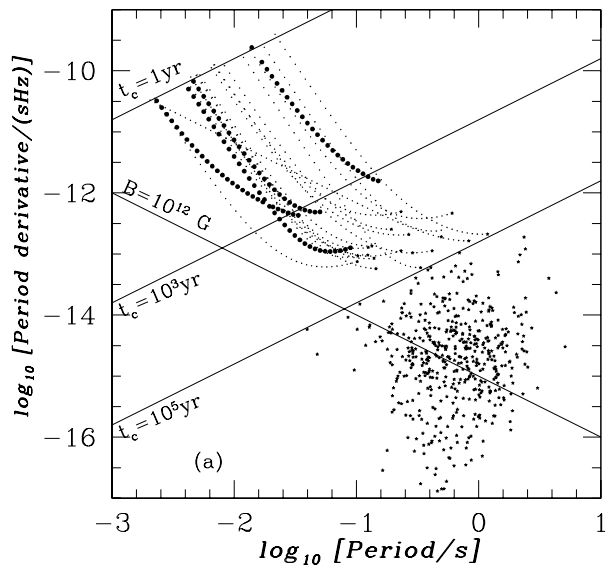, width=0.95\hsize}
\caption{Evolutionary trajectories for 17 middle aged pulsars evolved backwards to 
$t_{dyn}=1~\rm year$. The four youngest pulsars (Crab, PSR~B1509--58, PSR~B0540--69 
and Vela) are denoted by the large dots.}
\label{tr_mid}
\end{figure}

Although exact tracks cannot be computed for older pulsars, their known timing 
parameters not constraining $\{ g, r, s\}$ sufficiently, we evolved backwards 17 
pulsars with $t_{dyn} \leq 10^{5}~\rm years$, assuming $n=1.5$, in order to locate
the region of their likely initial conditions in the \pp diagram (Fig.~\ref{tr_mid}). 
We stopped the computation when $t_{dyn} = 1~\rm year$. We note that most of these 
pulsars arrive to higher $\dot{P}$ than all of the youngest pulsars, with the 
exception of PSR~B1509--58. In order to get to \pp positions more consistent with 
that of the Crab, these pulsars would need to have a braking index closer to 1. A 
prediction of the stationary multipole model is that $n\ga 1$ for the majority of 
the pulsar population.

\begin{figure}
\centering
\epsfig{file=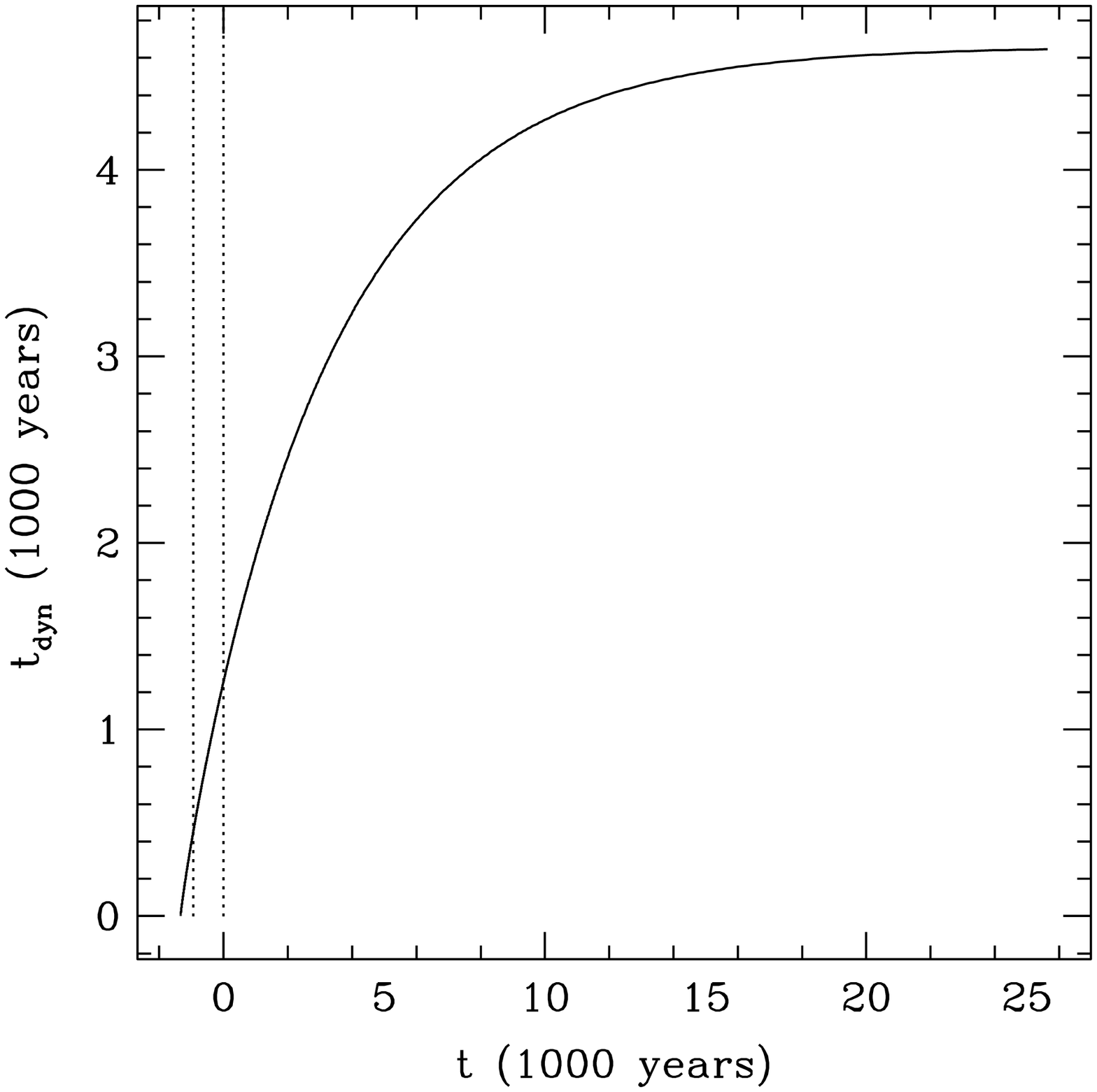, width=0.95\hsize}
\caption{Dynamical {\it versus} chronological time for the Crab pulsar. Note that 
$t_{dyn}\rightarrow 4.6\times 10^{3}$~yrs for ages greater than $\sim 20\times 
10^{3}$~years. 
As the model forces $\dot{\nu} \propto \nu$, as $n\rightarrow 1$, dynamical ages 
tend to constant values 
which, for the four pulsars studied, are below the values observed for other pulsars. 
The dotted lines mark the date of birth of the pulsar and today.}
\label{dynchron}
\end{figure}

A more detailed discussion of the stationary model can be found in 
Alvarez~(\cite{alvarez02}). Here we want to point out several inconsistencies of 
the model, which lead us to reject its non-time dependent version. As it can be 
seen in Fig.~\ref{tr1}, the evolutionary tracks of the youngest pulsars take them to 
a region of the \pp diagram where no pulsars have been found. Although a large 
$\dot{P}$ might difficult the detection of pulsars located there, the region 
corresponds to pulsars slower but more energetic than some known radio-pulsars, 
specifically those with same $\dot{P}$ but lower $P$. Pulsars in that region should 
be detectable. We believe these tracks are indicative of the inconsistency between 
the model and the observations.
A second discrepancy is the eventual dominance of the monopole term, which implies 
$\nu/\dot{\nu} \to\rm constant$ (Fig~\ref{dynchron}). As $n\to 1$ the dynamical time 
tends to a constant, $t_{dyn}\to 1/2s$, which values between $2.8\times 10^{3}$~years 
for PSR~B0540--69 and $13\times 10^{3}$~years for Vela. The dynamical ages of the
rest of the pulsars in the \pp diagram are much larger and cannot be reproduced by 
the model from the data of these young pulsars, a clear inconsistency between the 
predicted evolution of the young pulsars and the timing parameters of older pulsars. 

Through the study of the stationary multipole model proposed here we reach the general 
conclusion 
that {\em there is no time-independent evolutionary equation 
$\dot{\nu} = -f(\nu)$ able to reconcile braking indices with the \pp diagram}.
Note that including a $\dot{\nu}\propto\nu^{2}$ term\footnote{inconsistent with 
the antisymmetric condition.} cannot prevent the eventual dominance of the monopolar 
term, needed to explain the braking index of Vela (and probably also of
PSR~B0540--69).
The impossibility to have a time-independent expression means that the physical 
properties involved in the dynamics of pulsars must vary on timescales 
comparable to their life as active radio-pulsars. The most obvious form of time 
dependence is through the magnetic moment of the star, namely magnetic field decay 
or alignment between
the magnetic and rotational axes. We consider in the next 
section the time-dependent multipole model using physically expected 
time-dependences.

\section{The time-dependent multipole model} 
Modelling the distribution observed in the \pp diagram requires considering 
plausible time dependences in physical parameters involved like the magnetic 
dipole moment of the neutron star. Observational evidence in favor of magnetic 
field decay is discussed by Tauris \& Konar (\cite{Tauris}), who analyzed the 
relation between 
characteristic time against rotational energy of the pulsars in the Princeton 
Catalog~(Taylor et al. \cite{taylor}). Additional evidence comes from the measurements 
of their proper motions 
{and from the study of ultramagnetized neutron stars or magnetars.} 
Sang \& Chanmugam 
(\cite{sang}) argued that dynamical and kinetic ages of pulsars can 
only be 
reconciliated with decaying magnetic fields of the form $B(t)=B(0)/(1+t/t_c)$. 
{Further evidence of magnetic field decay is provided by anomalous X-ray pulsars 
(AXPs) interpreted as magnetars. Colpi et al. (\cite{colpi2000}) 
required the magnetic field in these neutron star related objects to decay as a 
power law on timescales of the order of $10^4$ years.}

We introduce time dependence in our model through the coefficients $\{r(t),s(t)\} 
\propto B(0)^{2}\psi(t/t_c)$, where $t_c$ is the characteristic time-scale for field
decay and $\psi$ a dimensionless function satisfying $\psi(0)=1$. Frequently used 
functional forms of $\psi$ are exponential or inverse linear. 
For simplicity, we will only consider future evolution for the youngest pulsars, so 
that
the quadrupolar coefficient $g(t)$ can be neglected hereafter\footnote{Neglecting 
$g$ also means that we do not require knowing $m$.}. 
We can then study evolutionary tracks on the \pp diagram writing 
equation~(\ref{model}) in terms of the period and its derivative as
\begin{equation}
\dot{P}=\left({r_0\over P}+s_0P\right)~\psi({t/t_c})\, .   \label{t-model}
\end{equation}
This equation describes the period evolution of pulsars for a decaying magnetic field. 
It can be integrated analytically, leading to:
\begin{equation}
 P^2(t)= \left({r_0/ s_0}+P^2_0\right)\exp \left[2s_0t_c~\Psi (t/t_c)\right] - 
 {r_0/ s_0} \, ,
\end{equation}
where $P_0 \equiv P(0)$ and $\Psi(x)=\int_{0}^{x} \psi(u)du$. 
If $~\psi\to 0~$ for 
$~t\gg t_{c}$, $\Psi$ will generally converge to a value 
$\Psi_\infty$ and the period will tend to a constant,
$P_\infty$.
This occurs because the spin-down mechanisms disappear as $\psi$ approaches zero. 
If one knows the present values of $\psi$ and $\dot{\psi}$, the initial parameters 
$r_0$ and $s_0$ can be derived from equation~(\ref{t-model}) and its time derivative:
\begin{eqnarray}
s_{0}  &=& {1\over 2\psi} \left({\dot{P}\over P}\right)  \left(3-n+{(-\dot{\psi})/
\psi \over \dot{P}/P}\right), \nonumber \\
r_{0}  &=& {P\dot{P}\over 2\psi}  \left(n-1-{(-\dot{\psi})/\psi \over \dot{P}/P}\right),
\end{eqnarray}
where $n$ is the present braking index and a decay-law means $\dot{\psi}< 0$. 
The decay timescale $t_c$ is taken here as a free parameter to fit the data, to be
determined by selecting physically meaningfull trajectories in the 
\pp diagram and $r_0>0$ and $s_0>0$. Note that $r_{0}$ and $s_{0}$ are both positive 
if
\begin{equation}
n - 1 > {(-\dot{\psi})/\psi\over \dot{P}/P} > n - 3 \, ,
\end{equation}
which relates $t_c$ with the dynamical time and requires \mbox{$n\geq 1$} {\em always}.
On the other hand, old pulsars can now have arbitrarily large braking indices, as $n$ can 
go to $\infty$ when $\psi/\dot{\psi}\rightarrow\mathrm{cte}$, $P\to P_{\infty}$ and 
$\dot{P}\to 0$. The largest loss of rotational energy occurs when $\dot{\psi}\to 0$, 
$\psi \to {\rm constant}$ i.e. when $n$ is close to one.

In the next sections we present evolutionary tracks for the cases of exponential and 
linear magnetic field decay.

\subsection{Exponential decay}
\begin{table}
\caption[]{Time-dependent model parameters and exponential 
magnetic field decay timescales consistent with $r_{0}\geq 0$ and $s_{0}\geq 0$.}
\label{tcs}
$$
\begin{array}{|l|cc|ccc|}
\hline
\noalign{\smallskip}
\mathrm{Pulsar} & r_{0} & s_{0} & \multicolumn{3}{c|}{t_{c}  ($yrs$)}\\
        &  ($Hz$^{-1}) & ($Hz$) & \mathrm{min}  & \mathrm{best} & \mathrm{max} \\
\noalign{\smallskip}
\hline
\mathrm{Crab}    & 1.0\cdot 10^{-14} & 4.0\cdot 10^{-12} & 3300 & 4.5\cdot 10^{4} & 8.0\cdot 10^{4} \\
{1509-58} & 2.1\cdot 10^{-13} & 1.8\cdot 10^{-12} & 3390 & 3.7\cdot 10^{4} & 1.5\cdot 10^{5} \\
{0540-69} & 1.0\cdot 10^{-14} & 6.8\cdot 10^{-12} & 6600 & 2.5\cdot 10^{4} & 5.0\cdot 10^{4} \\
\mathrm{Vela}    &
2.3\cdot 10^{-17} & 1.7\cdot 10^{-12} & 1.1\cdot 10^{5} & \ldots & 2.0\cdot 10^{5} \\
\noalign{\smallskip}
\hline
\end{array}
$$
\end{table}
\begin{figure} 
\centering
\epsfig{file=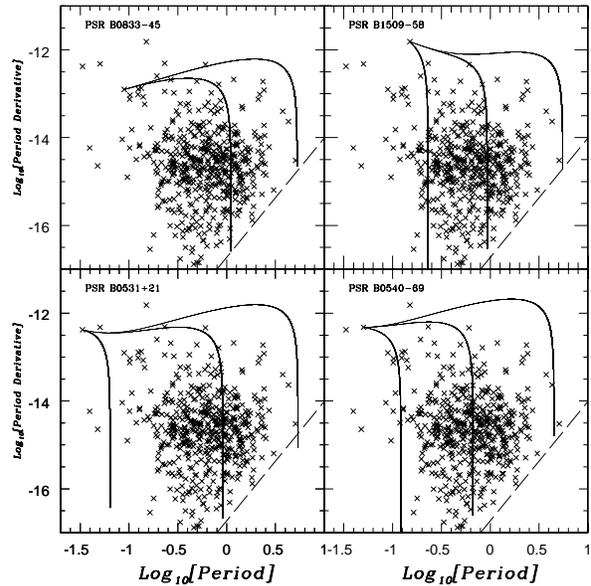, width=0.95\hsize}
\caption{Evolutionary trajectories using exponentially decaying magnetic fields for 
the Crab, PSR~B1509--58, PSR~B0540--69 and Vela. The three curves correspond to the 
lowest, best and largest $t_c$ consistent with observations. For Vela the best 
trajectory is that of minimum $t_c$, as shown in Table~\ref{tcs}. Trajectories were 
followed for 10$t_c$, 6$t_c$ and 5$t_c$ for the PSR~B0531+21, 5.5$t_c$ and 4$t_c$ for 
PSR~B0833--45, 100$t_c$, 5.5$t_c$ and 4$t_c$ for PSR~B1509--58 and 100$t_c$, 6$t_c$ 
and 4.8$t_c$ for PSR~B0540--69.}  
\label{tr_expo}
\end{figure}
Based on the observed deficit of pulsars with periods larger than 1 second, 
Ostriker \& Gunn (\cite{ostriker}) proposed a rapid decrease of pulsar radio 
luminosity as result of exponentially decaying magnetic fields,
\begin{equation}
B(t)=B_0\exp\left(-{t/ t_c}\right) \, ,
\end{equation}
a natural result of ohmic dissipation. In this case the model becomes
\begin{equation}
\dot{P}=e^{-2t/t_c}\left({r_0}/{P}+s_0P\right) \, ,
\end{equation}
and we have $\Psi_\infty = 1/2$.
For the youngest pulsars we obtain the trajectories shown in Figure~\ref{tr_expo}. 
Satisfactory trajectories are obtained for $t_c\approx 10^4$~yrs (Table~\ref{tcs}), 
which correspond to extremely rapid evolution, as the magnetic field becomes negligible 
in just $10~t_c$ and pulsars cross the death-line of the \pp diagram in $10^{4}$ 
to $10^{5}$ years. 
This timescales are too short to be consistent with estimated birth rates of one 
pulsar every 90 years (Brazier \& Johnston \cite{brazier99}) and supernova rates of one 
every 10-30 years 
(Van der Bergh \& Tammann \cite{vanderbergh}). We conclude that exponentially 
decaying magnetic fields cannot be used to properly fit the \pp diagram without 
contradicting present known pulsar and supernova rates.

\subsection{Inverse linear decay}
\begin{figure}
\centering
\epsfig{file=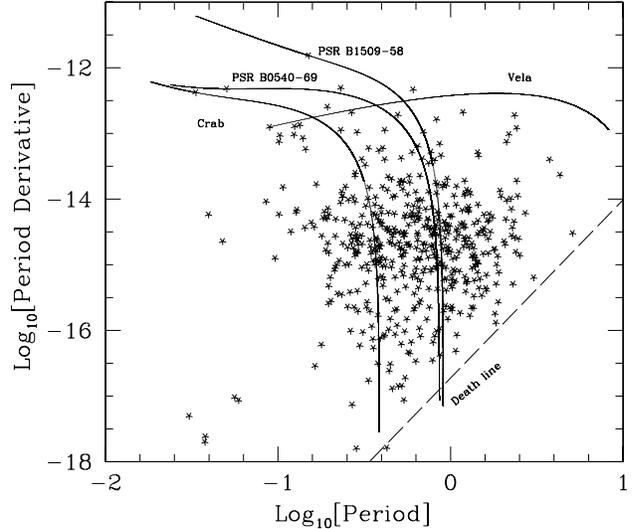, width=0.95\hsize}
\caption{Evolutionary trajectories for the Crab, PSR~B1509--58, PSR~B0540--69 and 
Vela, using an inverse linear decay-law for the magnetic field. The decay timescale 
used is $t_c = 10^{4}$ except for Vela where $t_{c} =1.04\times 10^{5}~\rm years$. 
The trajectories were run for $10^3~t_c$, except that of Vela which was run for 
$10t_c$.} 
\label{tr_inv}
\end{figure}

\begin{table}
\caption[]{Time-dependent model parameters and inverse linear 
magnetic field decay timescales consistent with $r_0\geq 0$ and $s_0\geq 0$.}
\label{tclinear}
$$
\begin{array}{|l|cc|ccc|}
\hline
\mathrm{Pulsar} & r_{0} & s_{0} & \multicolumn{3}{c|}{t_{c}  ($yrs$)}\\
        &  ($Hz$^{-1}) & ($Hz$) & \mathrm{min}  & \mathrm{best} & \mathrm{max} \\
\noalign{\smallskip}
\hline
\mathrm{Crab}    & 9.4\cdot 10^{-15} & 7.5\cdot 10^{-12} & 2090 & 1.0\cdot 10^{4} & 3.0\cdot 10^{4}\\
{1509-58} & 1.2\cdot 10^{-13} & 4.8\cdot 10^{-12} & 1830 & 1.0\cdot 10^{4} & 4.0\cdot 10^{4} \\
{0540-69} & 7.3\cdot 10^{-15} & 1.0\cdot 10^{-11} & 4980 & 1.0\cdot 10^{4} & 2.0\cdot 10^{4} \\
\mathrm{Vela}
   & 2.1\cdot 10^{-18} & 1.7\cdot 10^{-12} & 1.0\cdot 10^{5}  &  \ldots &  \ldots \\
\hline
\end{array}
$$
\end{table}
Sang \& Chanmugam (\cite{sang}) proposed an inverse linear decay law for the magnetic 
field of pulsars,
\begin{equation}
B(t) = {B(0) \over 1 + t/t_c} \, ,
\end{equation}
to reconcile dynamical and kinetic ages. Using this decay law in the model, relatively 
short timescales, \mbox{$t_{c} \sim 10^4$~years,} are needed to fit the evolutionary 
trajectories of the four pulsars considered with the \pp distribution, as shown
in Figure~\ref{tr_inv} for the Crab, PSR~B1509--58, PSR~B0540--69 and Vela. Except 
for the Vela pulsar, to be discussed below, trajectories can be fitted through 
the bulk of the pulsar population. Although the timescales needed to fit the data 
are similar to those for the case of exponential decay, linear decay is far more 
convenient as pulsars remain active for $10^{3}-10^{4}~t_c$, crossing the death line 
of radio emission in some $10^7$ years, in reasonable agreement with pulsar and 
supernova rates.

\begin{figure} 
\centering
\epsfig{file=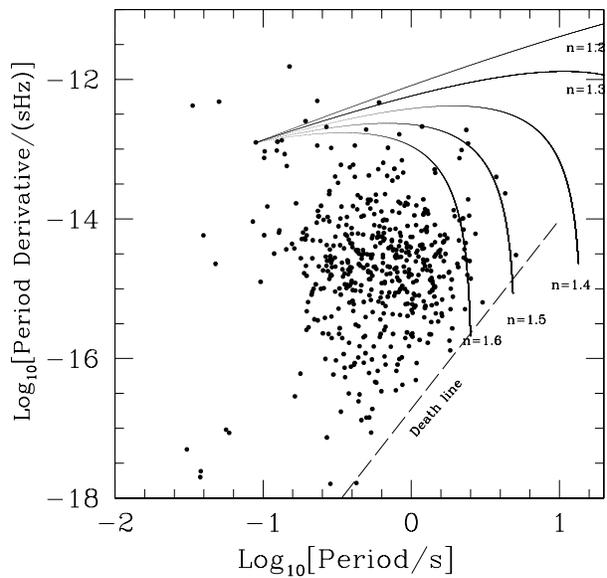, width=0.95\hsize}
\caption{Evolutionary trajectories for the Vela pulsar, considering different possible 
values of $n$. Trajectories using $n$ larger than $\sim 1.5$ are consistent with known
long-period
pulsars, while $n\leq 1.4$ is inconsistent with observations.}
\label{tr_vela}
\end{figure}
For the Vela pulsar, the minimum decay timescales consistent with the model, i.e. 
with positive $r_{0}$ and $s_{0}$, do not manage to turn its evolutionary tracks 
towards the bulk of pulsars. However, the uncertainty in the measurement of $n$ 
allows us to consider a range of values.
Evolutionary trajectories with $n\leq 1.4$ take Vela to a region of large periods 
and high period derivatives where no radio pulsars have been found. For $n=1.2$, Vela 
takes about $2.0\times 10^5$~yrs to slow down to $P\sim 10~\rm s$, and 3.6$\times 10^5$ 
years for $n=1.3$. Taking these numbers and considering one pulsar like Vela born every 
11000 years, we should expect to see between 18 and 27 pulsars in the same region of 
the \pp diagram occupied by Vela. Indices $1.5\leq n\leq
1.6$ can produce 
reasonable tracks (Fig.~\ref{tr_vela}), leading to longer periods than those of the 
other three young 
pulsars, but consistent 
with long-period pulsars present in the \pp diagram. It does not seem feasible to 
reconcile Vela trajectories with
those of the Crab, PSR~B1509--58 and PSR~B0540--69, 
which pass though the bulk of the pulsar population without impossing a braking index 
$n> 1.6$, inconsistent (at least at the 1$\sigma$ level) with observations. So, even 
though the model can fit the data, the evidence points to the Vela pulsar
as a 
particular type of pulsar. Pulsar evolution might distinguish between Crab-like 
pulsars, with magnetic-fields decaying at periods $P_{\infty}\la 1~\rm s$,
and -less 
common- Vela-like pulsars, where the magnetic-field decay is slower and becomes 
spun-down to asymptotic periods between 1 and 10~seconds.

\section{Physical Spin-down Mechanisms}
The main feature of the model presented here is the inclusion of the monopolar term, 
required to derive reasonable evolutionary tracks considering the measured braking 
indices.
The other two terms of the model have been widely used in the past and are usually
interpreted in terms of electromagnetic or gravitational radiation losses. An 
inclined rotating magnetic dipole in vacuum produces a dipolar term given by
\begin{equation}
\dot{E}_{dip}=-{2 \mu_{\perp}^{2}  \over 3c^3}~\Omega^4 \, ,  \label{dipole}
\end{equation}
where $\Omega = 2\pi\nu$ is the angular frequency and $\mu_{\perp}$ the component of 
magnetic dipole moment perpendicular to $\vec{\Omega}$.
Pacini (\cite{pacini67}) was the first to equate $\dot{E}_{dip} = I\Omega\dot{\Omega}$, 
introducing the basic equation of pulsar dynamics. A quadrupolar term can be 
associated either to the next electromagnetic moments or to the lowest gravitational 
radiation multipole moment:
\begin{equation}
\dot{E}_{quad}=-{32 G\over 5c^5}~I^2e^2~\Omega^6 \, , \label{quadrupole}
\end{equation}
where $Ie$ the product of the moment of inertia and  equatorial ellipticity of the 
star.
The fact that observed indices
are close to 3 indicates that gravitational radiation is not the dominant form of
energy loss in pulsars at present, although it might be important in the very early 
stages of pulsar evolution.

Directly equating the Larmor equation to rotational losses cannot lead to the 
monopolar term proposed in equation~(\ref{model}). The Larmor equation arises from
considering the energy flux of an electromagnetic wave, $\vec{S} \propto 
|\vec{E}_{rad}|^{2} \hat{k}$, with the radiation field $\vec{E}_{rad}$  
proportional to the acceleration of the charged particles, scaling therefore as 
$\Omega^{n}$ with $n$ even and $\geq 4$. Even in the framework of the
models mentioned in Section~2, it is unlikely that radiative emission processes
can give a different dependence on $\Omega$.

On the other hand, there is clear evidence that particle acceleration and massive
winds
are important energy loss processes in pulsars. This evidence includes the 
short 
lifetimes of relativistic electrons in the Crab Nebula and the need of a 
source to account for its
luminosity (as pointed out first by Oort and Walraven~\cite{oort56}), 
observational changes directly observed in the vicinity
of the Crab pulsar (Hester et al. \cite{hester}) and the X-ray images of the Crab, 
PSR~B1509--58 and the Vela pulsar (Weisskopf et al. \cite{weisskopf}, 
Gaensler et al.~\cite{gaensler}, and Helfand et al.~\cite{helfand} respectively). 
The monopolar term cannot be of radiative 
origin, so we propose here its relation with particle and/or mass loss processes. 
We note we cannot
discard a $\dot{E}\propto\Omega^{3}$ term but this term would be neither necessary nor 
sufficient to explain the data; therefore we avoid introducing it and assumed the 
antisymmetric condition for simplicity. Neglecting the $\dot{E}\propto\Omega^{3}$ 
term 
(equivalent to $\dot{\nu}\propto\nu^{2}$) we put forward the issue of 
whether particle acceleration and/or mass-loss follow an antisymmetric spin-down 
equation, like the Larmor equation does.

A monopolar spin-down term arises from any energy loss mechanism with a $\dot{E} 
\propto \Omega^{2}$ dependence, which can be expressed as
\begin{equation}
\dot{E}_{mon} = -\beta \left({\mu_{\perp}^{2} \over c^{3}}\,\Omega^{4} \right) 
\left({c/\Omega\over R_{*}}\right)^{2} \, ,
\label{mono}
\end{equation}
where $\beta$ is a dimensionless factor that can depend on time but not explicitly 
on $\Omega$. It is an open question whether energy-loss processes like particle 
acceleration and pulsar winds can provide this sort of expression and we can only
especulate about this point. Michel (\cite{michel}) estimated the torque 
exerted on a neutron star by a pulsar wind as 
\begin{equation}
T = ({\pi \Phi^{2} / 4c})~\Omega ,
\end{equation}
where $\Phi$ is the magnetic flux carried away by the wind. Assuming $\Phi$ can be 
constant, a pulsar wind would lead to a monopolar term in the energy loss equation. 
A detailed analysis of pulsar wind energetics has been given by Harding et al. 
(\cite{harding}), 
{who assumed an electric current flowing in the magnetosphere causing a magnetic 
torque on the star. The corresponding loss of rotational energy due to this torque is 
given by:}
\begin{equation}
\dot{E}_{wind} = {c \over 3}~B_{*}^{2}~R_{*}^{6}~\left({ 1 \over R_{open}^{2}~R_{LC}^{2}}\right)
\, ,
\end{equation}
{where $R_{LC} = c/\Omega$. $R_{open}$ is the distance from the center of the star 
to the magnetosphere region at which the magnetic field lines become open due to the 
pulsar wind.} For $R_{open}$ independent of $\Omega$, this implies $\dot{E}_{wind} 
\propto \Omega^{2}$, i.e., a monopolar-like term.

\section{Conclusions}
A multipole equation for pulsar evolution can be derived naturally as the Taylor 
expansion of the general equation $\dot{\nu} = -f(\nu,t)$. While it might give 
insight into short
term evolution of pulsars, the time independent version of the 
multipole equation is inconsistent with observations, indicating that no equation of 
the form $\dot{\nu} = -f(\nu)$ is
able to fully model pulsar spin-down evolution. 
A time dependent multipole equation considering
exponentially decaying magnetic-fields 
cannot reproduce the data without shortening the active life of pulsars below 
$10^{5}$~years. Introducing an inverse linear magnetic field 
decay with timescales of the order of $10^4$ years gives good agreement with observations 
and allows pulsars to remain active about $10^7$ years, in agreement with pulsar and supernova 
birth rates. The model is consistent with observations of the Vela pulsar only if
its braking index is $\ga 1.5$. The different trajectories of this pulsar relative to those of the
Crab, PSR~B1509--58 and PSR~B0540--69 suggest the existence two types of pulsars: 
those with relatively
short decay timescales which evolve to periods $P_{\infty} \la 1~\rm s$ and Vela-like
pulsars, with more persistent magnetic fields evolving to the larger period pulsars observed
in the \pp diagram.
We believe the need to introduce a monopolar term in the spin-down
equation is an indirect observational evidence that particle acceleration and/or mass-loss
processes must follow $\dot{E} \propto \Omega^{2}$. 

\acknowledgements 
C\'esar Alvarez thanks Dany Page for his valuable comments during the preparation of 
his Ph-D thesis. C\'esar Alvarez also thanks CONACyT for the Ph.~D. support grant 
received (reference number 86472).



\end{document}